\documentclass[12pt,aps,prd,preprint,tightenlines,superscriptaddress,
   showpacs,nofootinbib]{revtex4}
\usepackage{hyperref}
\usepackage{graphicx}
\usepackage{epsfig}
\usepackage{amssymb}
\usepackage{amsmath, amsfonts,euscript,bbm}
\usepackage{ifthen}
\newcommand{\roughly}[1]{\mathrel{\raise.3ex\hbox{$#1$\kern-0.85em
\lower1ex\hbox{$\sim$}}}}

\def\be{\begin{equation}}
\def\beq\begin{equation}
\def\ee{\end{equation}}
\def\bea{\begin{eqnarray}}
\def\eea{\end{eqnarray}}

\def\eqref#1{(\ref{#1})}

\def\UV{{\scriptscriptstyle U \kern-.1emV}}
\def\IR{{\scriptscriptstyle I\kern-.18em R}}

\def\gsim{\ \rlap{\raise 3pt \hbox{$>$}}{\lower 3pt \hbox{$\sim$}}\ }
\def\lsim{\ \rlap{\raise 3pt \hbox{$<$}}{\lower 3pt \hbox{$\sim$}}\ }

\begin{document}

\thispagestyle{empty}

\preprint{UCI-HEP-TR-2012-15}
\preprint{YITP-SB-12-36}

\title{Strange Couplings to the Higgs  \vspace*{0.5cm}
}

\author{Yanzhi Meng}

\affiliation{Department of Physics and Astronomy,
University of California, Irvine, CA 92697, USA \vspace*{0.5cm}}

\author{Ze'ev Surujon}

\affiliation{Department of Physics and Astronomy,
University of California, Irvine, CA 92697, USA \vspace*{0.5cm}}

\affiliation{C. N. Yang Institute for Theoretical Physics, 
Stony Brook University, Stony Brook, NY 11794, USA \vspace*{0.5cm}}

\author{Arvind Rajaraman}

\affiliation{Department of Physics and Astronomy,
University of California, Irvine, CA 92697, USA \vspace*{0.5cm}}

\author{Tim M.P. Tait}

%\vspace{2mm}
\affiliation{Department of Physics and Astronomy,
University of California, Irvine, CA 92697, USA \vspace*{0.5cm}}

\date{\today }

\pacs{12.60.-i, 95.35.+d, 14.80.-j}

\begin{abstract}
We explore the coupling of the strange quark to the state of mass close to 126 GeV
recently observed by the ATLAS and CMS experiments at the LHC.   
An enhanced coupling relative to the expectations for a SM Higgs has the effect of 
increasing both the inclusive production cross section and the partial 
decay width into jets.
For very large modifications, the latter dominates and the net rate into non-jet
decay modes such as diphotons is suppressed,
with the result that one can use observations of the diphoton decay mode to 
place an {\em upper} limit on the strange quark coupling.
We find that the current observations of the diphoton decay mode imply that the coupling of the new resonance
to strange quarks can be at most $\sim 50$ 
times the SM expectation at the $95\%$
C.L., if one assumes at most a ${\cal O}(1)$ modification of the coupling to gluons.
\end{abstract}

\maketitle

\section{Introduction}

For the first time since the discovery of the top quark in the 1990s, a new fundamental particle has been
discovered.  Data from both ATLAS \cite{:2012gk} and CMS \cite{:2012gu}, 
collected during the 2011 and 2012 LHC runs indicates the
presence of a new neutral boson with mass around 126 GeV and whose properties are strongly suggestive
of the long-sought Standard Model (SM) Higgs boson.
To establish this identification,
it is vital to measure the couplings of this resonance to the rest of the SM particles.
The LHC experiments have already made remarkable progress on this front,
with published  measurements of the branching ratios to gauge bosons, as well as to taus.  In fact, a number of
theoretical attempts to process the relatively limited number of experimental results into determinations of couplings
have recently appeared \cite{Low:2012rj,Giardino:2012dp,Carmi:2012in,Bonnet:2012nm,Plehn:2012iz,Djouadi:2012he}.

One sector that is particularly challenging to establish are the couplings of the light quarks (up, down, and strange).
Indeed, in the SM, these couplings are all very tiny, leading to no expected
visible phenomena at colliders.  In addition, light
quarks manifest as ``unflavored" jets of hadrons in detectors, and are very difficult to distinguish from each other
as well as from gluons.  If the newly discovered resonance is not precisely the SM Higgs, one could imagine that
(with some coincidences such that the rate works out to be approximately SM-like) 
it is actually being produced by some combination of the light quarks in addition
to by gluon fusion.  While perhaps not the most likely scenario, it is worthwhile to explore how we can be sure
that such a coincidence is not obscuring our understanding of the new resonance.  Excluding the more exotic
possibilities is an important step which is necessary to put the detailed program of measuring the Higgs couplings 
in the framework of more aggressive assumptions (such as the studies mentioned
above)
on more solid footing.

We choose to focus our attention on the coupling to the strange ($s$) quark.  
This is not because the strange is particularly
more likely to be responsible for the kind of conspiracy that we have in mind than the up ($u$) or down ($d$) quarks.
Rather, the strange quark is an interesting proxy for any light quark's coupling to the new resonance, and since the
strange has somewhat smaller, sea parton distribution functions (PDFs) whose shape is more reminiscent of the
gluon than the valence $u$ or $d$ quarks, it would be more likely to exploit kinematic degeneracy 
with production by gluon fusion.  Finally, the strange quark is expected in the SM to couple more strongly to the Higgs
by about an order of magnitude compared to $u$ or $d$.  Thus, if there is something peculiar about the light quark couplings,
it {\em could} reasonably reveal itself more readily in the strange coupling.

A much enhanced coupling to strange leads to several modifications to the 
properties of a Higgs-like particle.  First, it
opens a new contribution to the 
(typically unobservable at the LHC \cite{Berger:2002vs}) decay mode into jets
which if sufficiently enhanced can compete with other interesting decays, such as 
into photons or weak bosons.  Just as it has been noted that a SM Higgs of mass 
125 GeV provides many interesting
decay modes of comparable branching fraction to study, a new or subtly modified 
decay mode can more effectively hide at such a mass since several precise 
measurements are required to account for all of the decay modes.
Second, it enhances the production of the boson through
strange quark fusion, which is expected to be negligible in the case of the SM
Higgs.  If the coupling to gluons is also modified commensurately, 
it may be that the rate of production alone does not provide enough information
to isolate the two couplings.  We will allow for both couplings to gluon and to the
strange quark to vary, in order to explore
how rigorously one can determine both couplings in light of
this potential degeneracy.

Some interesting studies in a similar spirit (but asking different questions) 
can be found in 
Refs.~\cite{Carpenter:2012zr,Stolarski:2012he,Coleppa:2012eh,Bolognesi:2012mm,Boughezal:2012tz,Alves:2012fb}.

\section{Couplings}

We study an effective field theory consisting of the SM fermions and vector bosons, 
and the newly observed resonance, which we denote as $\phi$.  It is very useful
to contrast the couplings of $\phi$ with those of the SM Higgs at the same mass,
which we denote as $H$.

\subsection{Higgs Couplings}

We begin by reviewing the SM $H$ couplings to quarks and gluons (which is
completely dominated in the SM by the contribution from the top loop).
The SM quark interactions with Higgs are given by,
\bea
\mathcal{L}_{q\bar{q}H}
& = & 
-\sqrt{2} ~ \frac{m_q}{v} ~\sum_q ~\bar{q} q H ,
\eea
where $m_q$ is the quark mass and $v \simeq 246$~GeV is the Higgs
vacuum expectation value (VEV).
These couplings generate an effective coupling to gluons of the form \cite{djouadi}
\begin{equation}
\mathcal{L}  = -\frac{\alpha_s}{12 \pi v}  
\left(\sum_{q} ~ I_q \right)
H ~ G^{\mu\nu}_a G_{\mu\nu}^a 
\equiv  -~\frac{g^{\rm SM}_{ggH}}{v}~
H ~ G^{\mu\nu}_a
G_{\mu\nu}^a
\end{equation}
where
$\alpha_s$ is the strong coupling,
$G_{a}^{\mu\nu}$ is the gluon field strength tensor,
and the factor $I_q$ is,
\begin{equation}
I_q = 3\int_0^1 dx \int_0^{1-x} dy ~ \frac{1-4 xy}{1-xy/(m_q^2/M_H^2)-i \epsilon}. \nonumber\\
\end{equation}
In the heavy quark mass limit 
$4 m_q^2 / m_H^2 \gg1$ (which is a good approximation for $m_H \simeq 125$~GeV 
compared to the top quark of mass $m_t \simeq 171$~GeV) we have 
$I_t = 1$, and the effective coupling in the SM
$g^{\rm SM}_{ggH} = -i \alpha_{s} / (12 \pi )$.
It will be useful for later convenience to note that for the quark masses 
$m_s \simeq 100$ MeV, $m_b \simeq 4.1$ GeV, and $m_c \simeq 1.0$ GeV,
we have $I_s=0.00018$, $I_b=0.053$, and $I_c = 0.0088$, respectively.

\subsection{$\phi$ Couplings}

We parameterize the $\phi$ couplings relative to their SM counter-parts
via parameters $\kappa_i$, where we are particularly interested in $i=s, t, g$.
The strange and top quark couplings, we write as:
\bea
\mathcal{L}_{s\bar{s}\phi} & = & -\kappa_s ~\sqrt{2}~
\frac{m_s}{v} ~\bar{s} s \phi \label{eq: ssh_coupling} ~,\\
\mathcal{L}_{t\bar{t}\phi}
&=& -\kappa_t ~ \sqrt{2}~\frac{m_t}{v} ~\bar{t} t \phi ~.
\eea
Similarly, the coupling to gluons is written as,
\begin{equation}\label{eq: ggh_coupling}
\mathcal{L}_{gg\phi} =- \kappa_g ~
\frac{g^{\rm SM}_{ggH}}{v} ~ \phi ~ G_{a}^{\mu\nu} G^{a}_{\mu\nu}~.
\end{equation}
Because the gluon coupling results primarily from a top loop, there is in general a connection between
$\kappa_g$ and $\kappa_t$.  In principle, for very large values of $\kappa_s$, it may also contribute, as will
any exotic colored particle with substantial coupling to the Higgs.  
In the absence of such colored particles and for $\kappa_s \lesssim 10^3$, one arrives at the relationship
$ \kappa_g \simeq \kappa_t$.  We derive results both assuming this relationship holds, and also for the 
case $\kappa_t$ is fixed to one, and allowing $\kappa_g$ to vary independently.

Though not of foremost interest here, large enhancements of $\kappa_t$ will 
also be reflected in an enhancement of
the associated production of $\phi$ with a pair of top quarks.  A recent CMS search for associated
production $p p \rightarrow t \bar t \phi$ 
(with $\phi \rightarrow b \bar{b}$) finds an upper limit on this process
of 8.5 times the SM expectation at $95\%$ confidence level 
(CL) \cite{CMS-tth}.  
Assuming a standard BR into $b \bar{b}$, this requires $\kappa_t \lesssim 3$.  While not an independent limit because
of the dependence on the coupling to bottom quarks, we will restrict ourselves to consideration of the region of
parameter space such that $\kappa_t < 3$.

\section{Decay widths, Branching ratios, and Production Rates}

Including the leading QCD correction, the partial decay width of the Higgs to $gg$ is \cite{djouadi}
\begin{equation}
\Gamma \left(H \rightarrow gg \right) = 
\frac{G_F \alpha_s^2 M_H^3}{36 \sqrt{2}\pi^3}\left[1+\left(\frac{95}{4}-\frac{7}{6}N_f\right)\frac{\alpha_s}{\pi}\right],
\end{equation}
where $N_f=5$ is the number of light quarks and the strong coupling constant $\alpha_s$ is understood to be
evaluated at $M_H$.
The partial decay width to $s \bar{s}$ is,  
\begin{equation}
\Gamma \left(H \rightarrow s \bar{s} \right)= \frac{3 G_F M_H}{4\sqrt{2} \pi}
~ \overline{m}_s^2(M_H) ~ \left[1+\Delta_{ss} + \Delta_H^2\right],
\end{equation}
where large logarithms are absorbed into the $\overline{\rm MS}$ strange mass 
$\overline{m}_s$ evaluated at $M_H$, $\overline{m}_s (125~{\rm GeV}) = 82.6$~MeV \cite{Fusaoka:1998}.
The QCD correction factors $\Delta_{ss}$ and $\Delta^2_H$ are modest in size, and 
may be found in Ref.~\cite{djouadi}.  The decay width into ``unflavored"
jets is given by the sum of the partial widths into gluons and into strange quarks.

The SM Higgs decay into photons is dominated by contributions from the $W$ boson and top quark
\cite{djouadi}, 
\begin{equation}
\Gamma \left(H \rightarrow \gamma\gamma \right) = \frac{G_{F} \alpha^2 M_H^3}
{128 \sqrt{2}\pi^3} \left|\sum_f N_c Q_f A_{1/2}(\tau_{f}) + A_1(\tau_W)\right|^2,
\end{equation}
where $\alpha$ is the electromagnetic coupling and
the form factors for spin-$\frac{1}{2}$ and spin-1 particles $A_{1,1/2}(\tau)$ are given in terms
of the ratio $\tau_i = M_H^2/(4M_i^2)$ (where $M_i$ is the mass of the intermediate particle in the loop) by
\begin{eqnarray}
A_{1/2}(\tau) &=& ~2[\tau + (\tau - 1) f(\tau)]\tau^{-2} ~,\nonumber\\
A_1(\tau) &=& -[2\tau^2 + 3 \tau + 3 (2\tau - 1) f(\tau)]\tau^{-2}~,
\end{eqnarray}
with the function $f(\tau)$ is defined as
\begin{equation}
f(\tau) = \left\{\begin{array}
{r@{\quad   \hspace{2 cm}  \quad}l}
\arcsin^2 \sqrt{\tau}     &\tau \le 1\\
-\frac{1}{4}\left[\log\frac{1+\sqrt{1-\tau^{-1}}}{1-\sqrt{1-\tau^{-1}}}
-i \pi \right]^2
&\tau > 1
\end{array}\right. ~.
\end{equation}
For a Higgs mass of 125 GeV, the total decay width is
$\Gamma_H = 4.07$ MeV and one has
${\rm BR}(H\rightarrow gg) \approx 8.57 \times 10^{-2}$ and
${\rm BR}(H\rightarrow s \bar{s}) \approx 5.15 \times 10^{-4}$ \cite{higg_cross_section,djouadi}.

Allowing for $\kappa_s$ and $\kappa_g = \kappa_t$ to deviate from unity, the partial widths for $\phi$ to decay into $gg$
and $s \bar{s}$ are modified in an obvious way,
\bea
\Gamma \left( \phi \rightarrow gg \right) &=& \kappa_g^2 ~ \Gamma \left( H \rightarrow gg \right) ~,\\
\Gamma \left( \phi \rightarrow s \bar{s} \right) &=& \kappa_s^2 ~ \Gamma \left( H \rightarrow s \bar{s} \right)~.
\eea
The implicitly modified coupling to top quarks, together with the modification to the strange coupling may also influence the
partial width into $\gamma \gamma$,
\begin{equation} \label{eq: modifiedgammagamma}
\Gamma \left(\phi \rightarrow \gamma\gamma \right) = \frac{G_{F} \alpha M_{\phi}^3}{128 \sqrt{2}\pi^3}
\left| 3 ~\kappa_g \left(\frac{2}{3}\right)^2 A_{1/2}(\tau_t)+ 3~\kappa_s
\left(-\frac{1}{3}\right)^2 A_{1/2}(\tau_s) + A_1(\tau_W)\right|^2,
\end{equation}
where we assume that there are no exotic charged particles with strong coupling to the Higgs making a relevant
contribution.  The full $\phi$ width can be expressed in terms of the SM Higgs width, $\kappa_s$, and $\kappa_g$ as,
\begin{equation}\label{eq: total_decay_width}
\Gamma_{\phi} = \Gamma_H~ \left[1 + (\kappa_g^2 - 1)~ {\rm BR}
(H\rightarrow gg) + (\kappa_s^2 - 1) ~ {\rm BR}(H\rightarrow s \bar{s}) \right]~,
\end{equation}
where we neglect the small contribution due to the modification in the partial width into photons.  One can already infer
a weak bound on $\kappa_s$ based on the fact that the distribution of $\gamma \gamma$ events associated with the
$\phi$ discovery is consistent with the experimental resolution on the diphoton invariant mass:  for $\kappa_s \gsim 500$
and $\kappa_g \sim 1$,
one would have $\Gamma_\phi \gsim 1$~GeV.

Modifications to the coupling to strange quarks and/or gluons also implies that the production rate
will be modified.  The dominant production process at the LHC for the SM Higgs is gluon fusion (GF), 
$gg \rightarrow H$. If
the resonance has large enough couplings to the strange quark, strange quark fusion (SF), $s \bar{s} \rightarrow H$
may also contribute relevantly.
The inclusive $\phi$ production rate can be expressed as
\begin{equation}\label{prodcrosssection}
\sigma(pp \rightarrow \phi) = \kappa_g^2~ \sigma_{\rm GF}(pp \rightarrow H) +
 \kappa_s^2 ~\sigma_{\rm SF}(pp \rightarrow H).
\end{equation}
where we use the SM gluon fusion rate $\sigma_{\rm GF}(pp \rightarrow H)$ at 
NNLO as given in Ref.~\cite{higg_cross_section}.
The strange fusion rate $\sigma_{\rm SF}(pp \rightarrow H)$ is computed at tree level using Madgraph 5
\cite{Alwall:2011uj}.  While higher order calculations of $s \bar{s} \rightarrow H$
are not available, the process of $b \bar{b} \rightarrow H$ may provide a reasonable guide that
corrections of order $+ 25\%$ are indicated for a mass of 125 GeV and LHC energies \cite{Harlander:2003ai}.

\begin{figure}[t]
%\label{fig: kg_var}
\begin{center}
\includegraphics[width=8.15 cm]{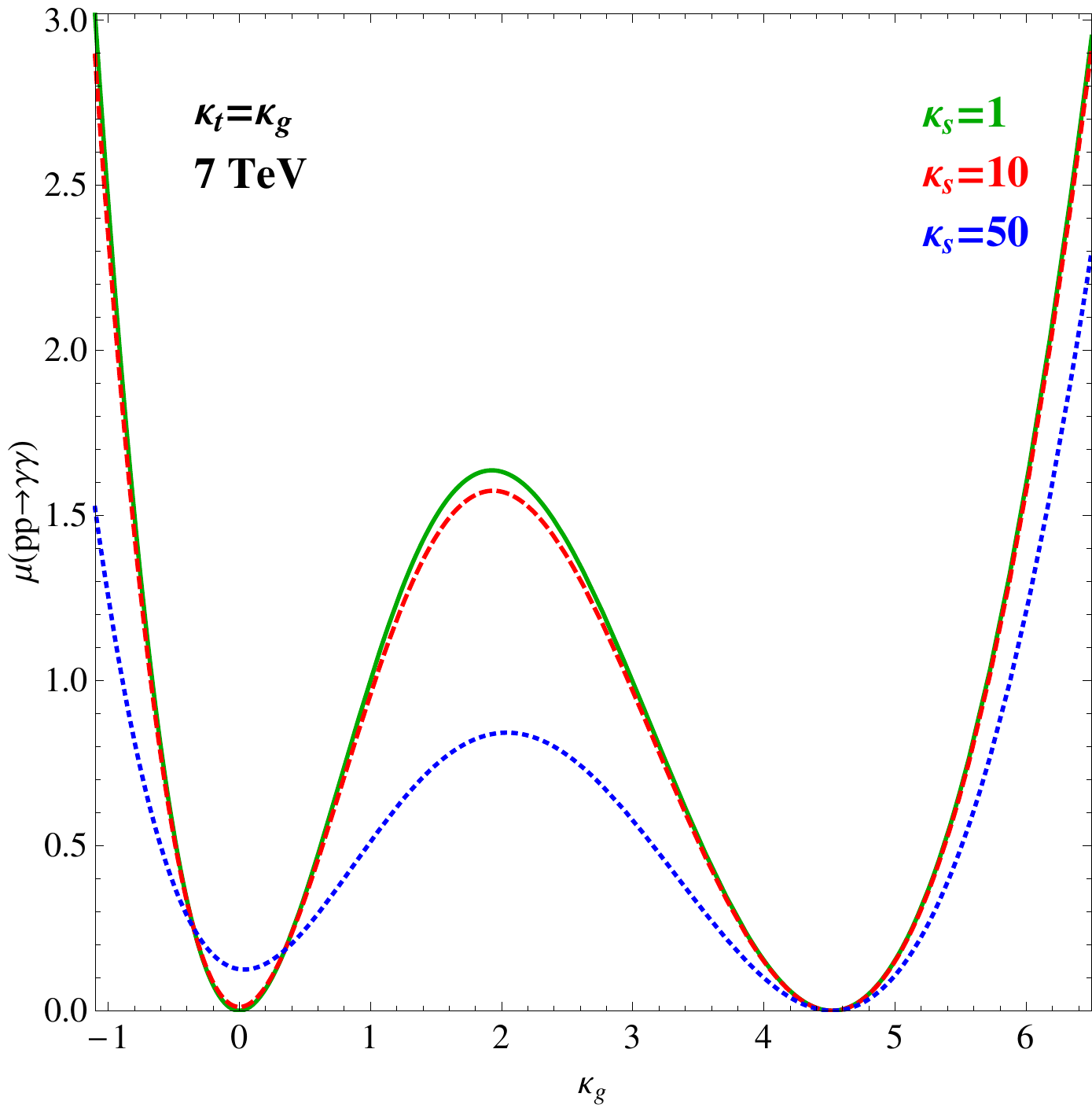}
\includegraphics[width=8.0 cm]{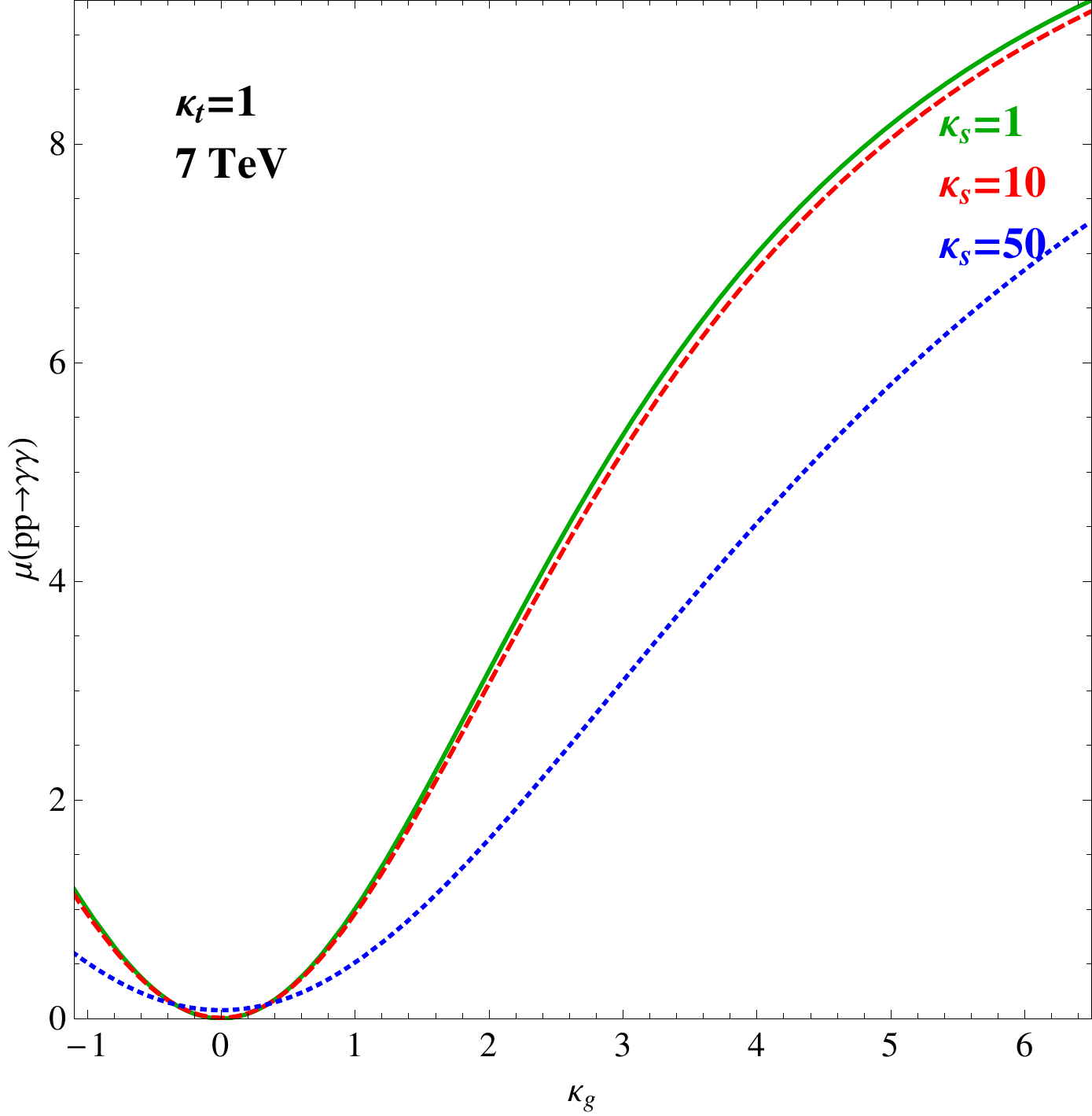}
\end{center}
\caption{Enhancement of the 
diphoton cross section ratio at the 7 TeV LHC as a function of 
$\kappa_g$, for $\kappa_s = $ 1 (green/top), 10 (red/middle), and 50 (blue/bottom), for
the case $\kappa_t = \kappa_g$ (left panel) and $\kappa_t = 1$ (right panel).}
\label{crossecfixedKs}
\end{figure}

The $\phi$ contribution to the diphoton cross section is
\begin{equation}\label{gammagamma}
\sigma(pp \rightarrow \phi \rightarrow \gamma \gamma) = \sigma(pp \rightarrow \phi)
\times \frac{\Gamma(\phi \rightarrow \gamma \gamma)}{\Gamma_{\phi}}~.
\end{equation}
Because increasing $\kappa_g$ increases both the production and $\Gamma_\phi$,
one cannot arbitrarily increase the production of diphoton events.  This is illustrated
in Figure~\ref{crossecfixedKs}, where we plot the ratio $\mu$ of
$\sigma(pp \rightarrow \phi \rightarrow \gamma \gamma)$ to the SM prediction at
the same mass, $\sigma(pp \rightarrow H \rightarrow \gamma \gamma)$,
at 7 TeV center of mass energy and for three representative choices of
$\kappa_s = 1, 10$, and 50.  For $\kappa_t = \kappa_g$ and modest (${\cal O}(1)$) deviations in
$\kappa_g$, the largest possible increase is about 1.63, which
occurs for $\kappa_g \simeq 2$ and SM-like coupling to the strange quark.  Larger increases are possible
only for more extreme modifications of $\kappa_g$ (in particular for the case of fixed $\kappa_t = 1$).

\section{Constraining $\kappa_s$ and $\kappa_g$}

The inclusive Higgs production rate relative to the SM prediction singles out
regions of the $\kappa_s$-$\kappa_g$ plane.  From Figure~\ref{crossecfixedKs},
one can see that there may be one or more (disjoint) regions which can realize
a given ratio.  The current determinations \cite{:2012gk,:2012gu},
\bea
\mu \equiv 
\frac{p p \rightarrow \phi \rightarrow \gamma \gamma}
{p p \rightarrow H \rightarrow \gamma \gamma}
& = & \left\{ \begin{array}{lr}
1.9 \pm 0.5 ~~~~~ & ~~~~~{\rm (ATLAS)} \\
1.56 \pm 0.43 ~~~~~ & ~~~~~{\rm (CMS)}
\end{array}
\right. ~,
\eea
are roughly consistent with the SM expectations, and may very well end
up settling close to one with more data.  For the moment, we consider two
reference
scenarios with a ratio of 1.0 (corresponding to a SM-like central value for the
inclusive production rate), and 1.63 (suggested by the current data).

\begin{figure}[t]
\begin{center}
\includegraphics[width=8.0 cm]{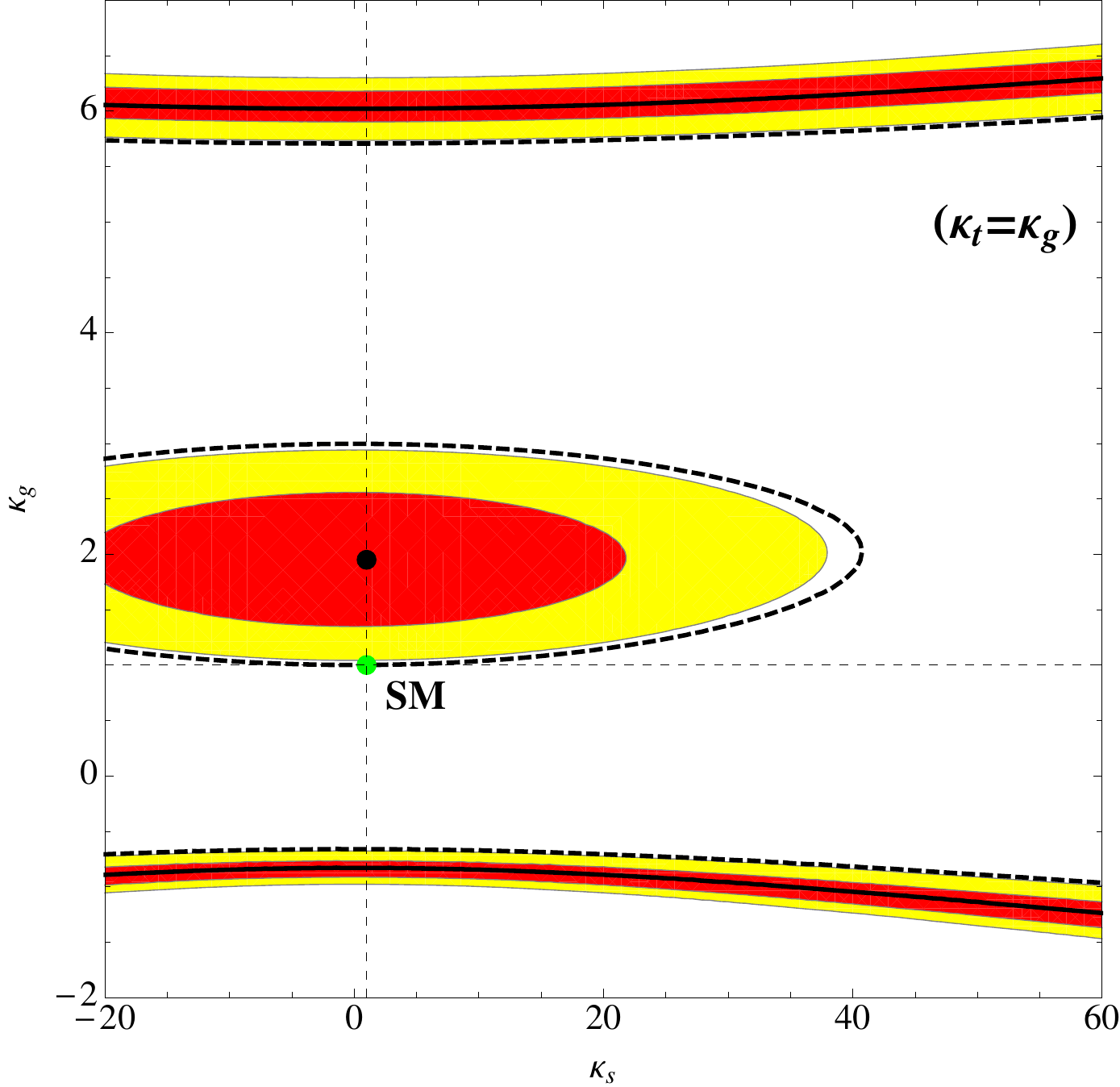}
\includegraphics[width=8.0 cm]{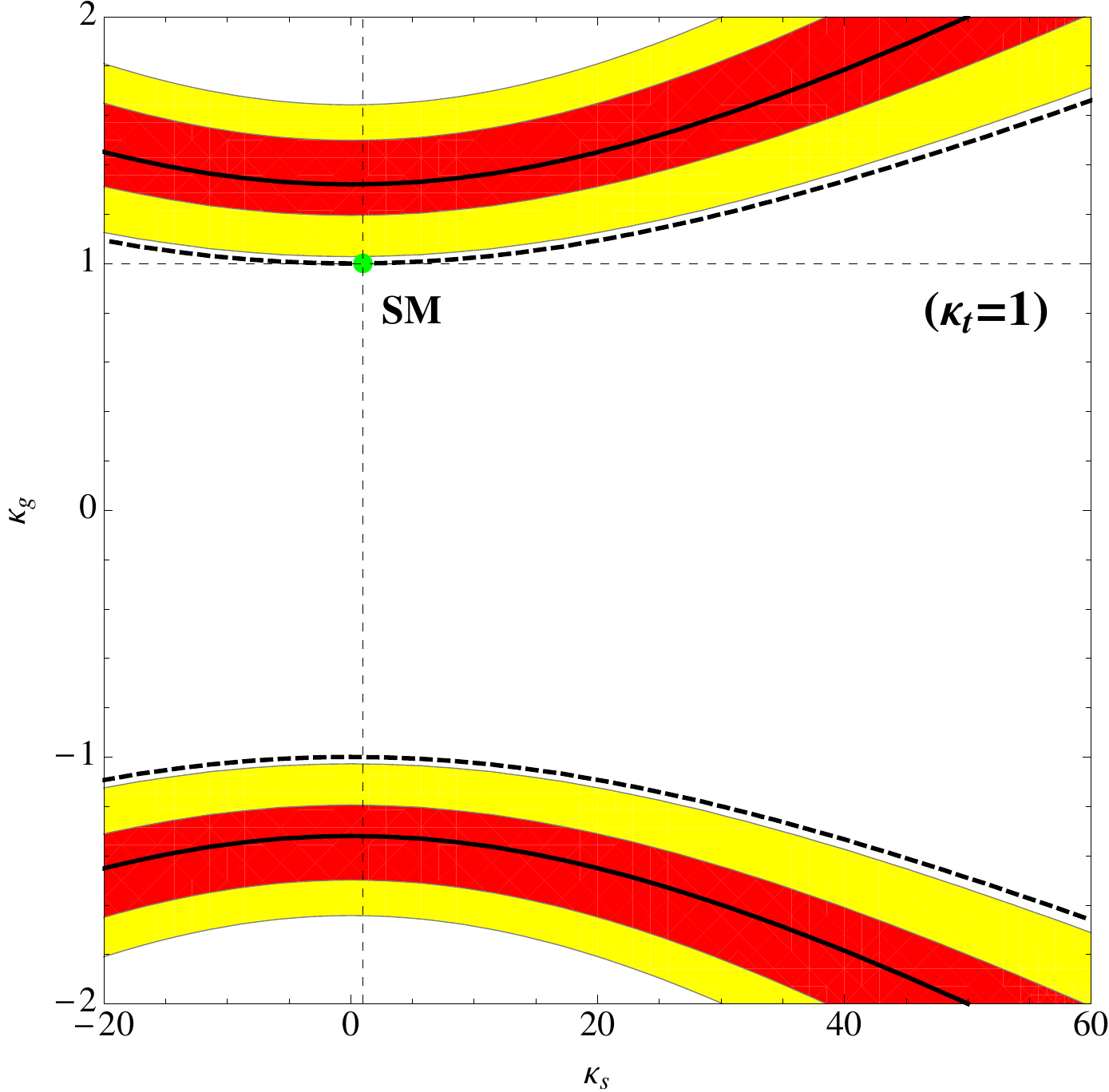}
\end{center}
\caption{$\kappa_g$ vs. $\kappa_s$ satisfying requirement of diphoton cross
section $\sigma (pp \rightarrow \phi \rightarrow \gamma\gamma) =1.63 \times
\sigma (pp \rightarrow H \rightarrow \gamma\gamma)$ (black solid curves and black dot),  
and
$ = 1.0 \times \sigma (pp \rightarrow H \rightarrow \gamma\gamma)$ (dashed curves). 
Also shown is the region consistent with the combined ATLAS and CMS
data at $1\sigma$ (shaded red region) and $2\sigma$ (shaded yellow region). SM is showed by the green dot}
\label{KgKs}
\end{figure}

In Figure~\ref{KgKs}, we show the curves of $\kappa_g, \kappa_s$ which result
in a constant $p p \rightarrow \phi \rightarrow \gamma \gamma$ cross
section of 1.63 or 1.0 times the SM rate, and the regions consistent with
the combined ATLAS and CMS measurements.  
Except for extremely modified $\kappa_g$, the curves for do not extend to arbitrarily large $\kappa_g$, $\kappa_s$,
because as illustrated in Figure~\ref{crossecfixedKs} the diphoton rate will
eventually be swamped by decays into jets.   Already, observation of the decay
into diphotons provides powerful information about $\kappa_s$ : for $\kappa_g = \kappa_t$, at $1\sigma$ the
data requires $\kappa_s \lesssim 25$ and at $2\sigma$, 
$\kappa_s \lesssim 60$.  Somewhat counter-intuitively, a smaller
value for the diphoton ratio permits {\em larger} modifications of the strange
coupling to the Higgs.  We can safely rule out the possibility that the observed
Higgs-like state is being produced solely by strange fusion as opposed to gluon
fusion.

%However, if the diphoton cross section turns out to be close to the SM expectation,
%then we would have to find another approach to constrain $\kappa_s$. We note that the total
%decay width is already known to be less than a GeV; this puts a weak constraint on $\kappa_s$.
%The  total decay width of the resonance as a function of $\kappa_g, \kappa_s$
%is shown in Fig.(\ref{width}); it can be seen that this sets $\kappa_s<500$.

%\\begin{figure}[h] 
%\label{fig: gamma_total}
%\begin{center}
%\includegraphics[width=10.0 cm]{gamma_total}
%\end{center}
%\caption{Variation of total decay width of LHC scalar with $\kappa_s$.}
%\label{width}
%\end{figure}

Going beyond the inclusive rate, one can imagine that more exclusive 
measurements could shed light on the $\kappa_s$-$\kappa_g$ degeneracy.
For example, the rate of additional hard jet radiation 
should be larger for gluon-initiated processes than for quarks, and the balance
of flavor content will likewise be different.  Both of these types of observables
may eventually come under control, but require large samples and
very precise theoretical understanding of Higgs production to be used reliably.

%\begin{figure}[t] 
%\begin{center}
%\includegraphics[width=12.0 cm]{sigma_photonphoton}
%\end{center}
%\caption{Diphoton cross section $pp\rightarrow \phi \rightarrow \gamma\gamma$
%as a function of energy, for $\kappa_s = 1$ and $\kappa_g$ adjusted such that
%the rate at 8 TeV is $1$ or $1.6$ (black/upper and
%red/lower curves, respectively) times the SM prediction. }
%\label{sigma_photonphoton}
%\end{figure}

\begin{figure}[t]
\begin{center}
\includegraphics[width=8.0 cm]{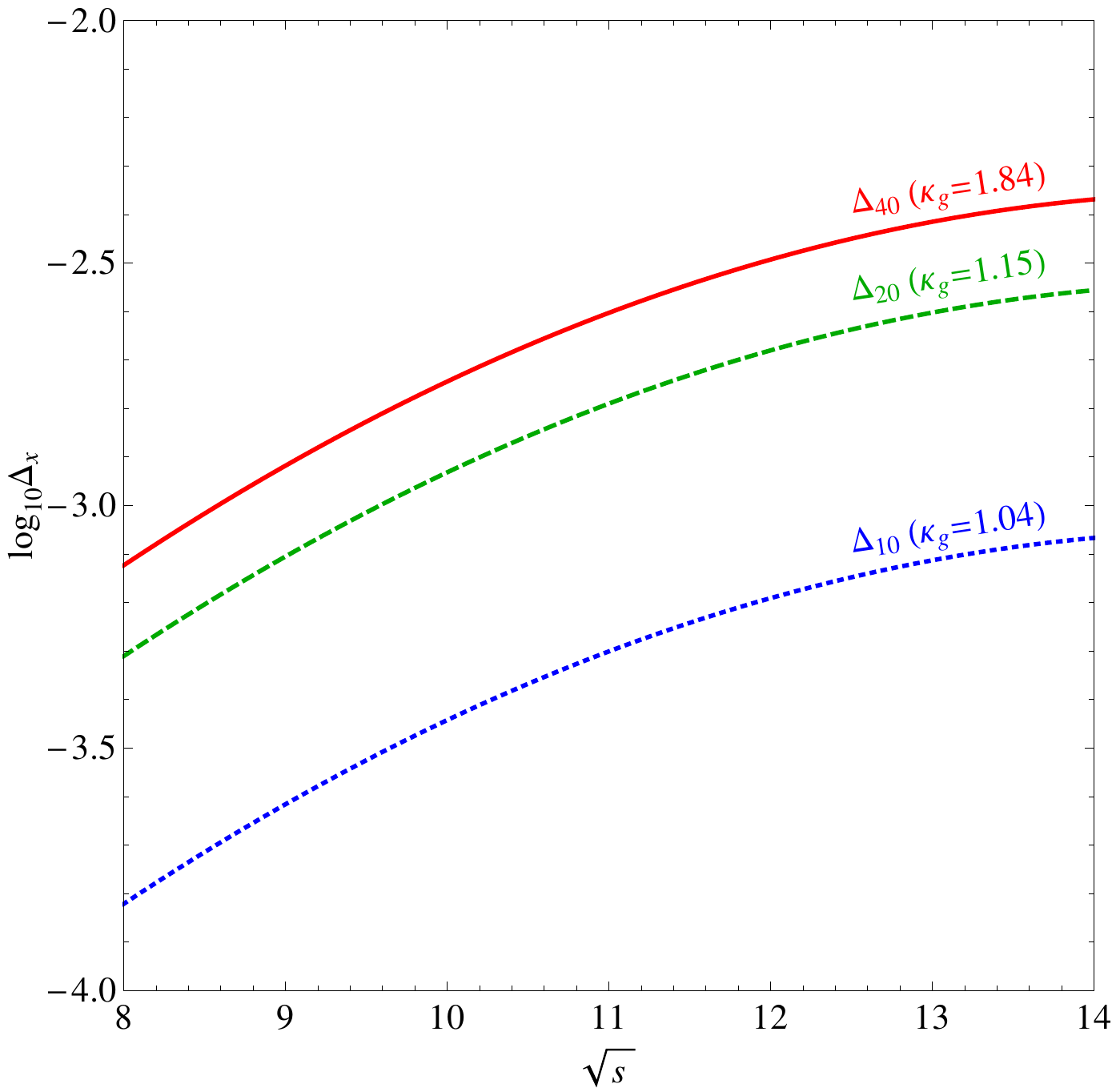}
\end{center}
\caption{$\log_{10}$ of the fractional difference in the rate of $p p \rightarrow \phi \rightarrow \gamma \gamma$ between
a model with $\kappa_s = 1$ and $\kappa_s = 50~(10)$, indicated by the upper (lower) curve, with
$\kappa_g$ adjusted such that the over-all rate at $\sqrt{s} = 7$~TeV is SM-like. }
\label{deltasigma}
\end{figure}

A simple inclusive feature that one can exploit is the fact that the LHC provides data at multiple collider
energies, and look at the inclusive rate and how it changes with energy.
Because the strange and gluon PDFs have slightly different slopes, the production
rates for each case exhibit a small difference in their energy dependence.  Unfortunately, the dependence
turns out to be rather weak.  If one defines the difference between two hypotheses for $\kappa_s$ to be
$\Delta_x$,
\begin{equation}
\Delta_{x} = \log_{10}~\frac{\sigma_{\gamma\gamma}(\kappa_s = 1)-\sigma_{\gamma\gamma}(\kappa_s = x)}{\sigma_{\gamma\gamma}(\kappa_s = 1)},
\end{equation}
where the subscript 
$x$ refers to the relative difference between the cross section at $\kappa_s = x$ with that for which $\kappa_s = 1$.
In Figure~\ref{deltasigma}, we plot $\Delta_{10}$ and $\Delta_{40}$
as a function of energy, with $\kappa_g$ adjusted in each case such that the cross section at $\sqrt{s} = 7$~TeV is SM-like.  
Even for extreme cases, the difference remains below
$1\%$, well within uncertainties
from theoretical sources (as well as any reasonable expectation for statistics).

\section{Outlook}

In this article, we have examined the dependence on the inferred properties of the newly discovered boson with mass
around 125 GeV on assumptions concerning its coupling to the strange quark.  Somewhat counter-intuitively, we find that
the largest impact of an enhanced strange coupling is to increase the total width, and thus depress the branching ratio
into $\gamma \gamma$.  As a result, a measurement of the diphoton rate which is enhanced compared to the SM
expectation actually provides a stricter upper bound on the strange coupling than a rate which is more consistent with the
SM expectation.

We find that the current observation of decays into diphotons at an enhanced rate places a rather strong bound that the
strange coupling be enhanced by no more than $\kappa_s \leq 25~(50)$ at 
the $1\sigma$ ($2\sigma$) level, if one assumes ${\cal O}(1)$ modification to the coupling to gluons.  A rate into diphotons
which was closer to the SM expectation would have resulted in a weaker bound.  We examined the energy dependence of
the diphoton cross section, and found that the energy dependence
of scenarios which saturate the $\kappa_s$ limit is typically too small
to resolve given the intrinsic theoretical uncertainties.

As we unravel the properties of the newly discovered boson, it may well turn out to be consistent with a vanilla, SM Higgs.
However, it is important to explore alternatives, even those that may seem unlikely, in order to avoid the pitfall of scenarios
that look superficially SM-like, but are in fact exotic.  In the current work, we found, somewhat to our surprise, that
there is not a large slop in the strange quark couplings to the Higgs, and one can place a relatively strong bound given some
mild assumptions.  Of course, it remains to be seen whether the picture will ultimately clarify into SM expectations, or something
a different, and the path is assured to be interesting and illuminating.

\acknowledgements

ZS and 
TMPT acknowledge the Aspen Center for Physics (supported by the NSF grant PHY-1066293) 
where part of this work was completed.
and the theory group of SLAC National Lab for their support of his many visits.
The work of AR and TMPT is supported in part by NSF grants PHY-0970173 and PHY-0970171.

%****************************************************************************
%
%\begin{center}\label{table: madgraph_calculation}
%\begin{tabular}{c|ccccc}
%$\sigma(pp\rightarrow H)$(pb) & 7 TeV   &   8 TeV  &10 TeV &12 TeV & 14 TeV    \\\hline
%$gg\rightarrow H$  & 15.32 & 19.52 & 28.733 & 38.92 & 49.85 \\
%$s \bar{s}\rightarrow H$  & $6.1  \times 10^{-4}$ & $7.610  \times 10^{-4}$ & $1.086 \times 10^{-3}$ & $1.435  \times 10^{-3}$ & $1.803  \times 10^{-3}$
%\end{tabular}
%\end{center}
%
%****************************************************************************

 \end{document}